\documentclass[twocolumn,english]{IEEEtran}
\usepackage[T1]{fontenc}
\usepackage{babel}
\usepackage{amsthm}
\usepackage{graphicx}
\usepackage[unicode=true,
 bookmarks=true,bookmarksnumbered=true,bookmarksopen=true,bookmarksopenlevel=1,
 breaklinks=false,pdfborder={0 0 0},backref=false,colorlinks=false]
 {hyperref}
\hypersetup{pdftitle={Metcalfe's Law Revisited},
 pdfauthor={Dmitri Nosovicki},
 pdfsubject={Network Theory},
 pdfkeywords={computer, network, information, economics},
 pdfpagelayout=OneColumn, pdfnewwindow=true, pdfstartview=XYZ, plainpages=false}

\makeatletter
\theoremstyle{plain}
\newtheorem{thm}{\protect\theoremname}
\theoremstyle{definition}
\newtheorem{defn}[thm]{\protect\definitionname}
\theoremstyle{plain}
\newtheorem{lem}[thm]{\protect\lemmaname}
\theoremstyle{plain}
\newtheorem{cor}[thm]{\protect\corollaryname}

\ifCLASSOPTIONcompsoc
\usepackage[caption=false,font=normalsize,labelfont=sf,textfont=sf]{subfig}
\else
\usepackage[caption=false,font=footnotesize]{subfig}
\fi
\usepackage{pgfplots}
\usepackage{gnuplottex}

\makeatother

\providecommand{\corollaryname}{Corollary}
\providecommand{\definitionname}{Definition}
\providecommand{\lemmaname}{Lemma}
\providecommand{\theoremname}{Theorem}

\begin{document}

\title{Metcalfe's Law Revisited}

\author{Dmitri Nosovicki%
\thanks{e-mail: \protect\href{http://nosovicki@ieee.org}{nosovicki@ieee.org}.%
}}
\maketitle
\begin{abstract}
Rudimentary mathematical analysis of simple network models suggests
bandwidth-independent saturation of network growth dynamics, as well
as hints at linear decrease in information density of the data. However
it strongly confirms Metcalfe's law as a measure of network utility
and suggests it can play an important role in network calculations.\end{abstract}

\section{Introduction}

\IEEEPARstart{M}{etcalfe's} law relates to communications networks;
it states that the value of a network is proportional to square of
its size. Not long ago, a group of authors\cite{Briscoe06c:nlogn}
challenged quadratic dependence. That split community into believers
and deniers of Metcalfe's law, and their claim was recently challenged
by statistical data\cite{zhang2015tencent}. However no attempts were
made to establish the truth mathematically, perhaps due to difficulties
with obtaining mathematical definition of ``value''. This paper
establishes a notion of value and analyses two conflicting models
of network. First, traditional model, fails to manifest Metcalfe's
law. Another model, that observes network in a wider context, both
confirms Metcalfe's law and shows its upper boundary.

\section{Network Value}

In lines of Von Neumann\textendash{}Morgenstern utility theorem\cite{von2007theory},
\begin{defn}
Utility of a system is a probability-weighted sum of its value for
all possible events:
\[
U=\sum_{i}^{\infty}\phi(K^{n},e_{i})P(e_{i})
\]

where $U_{S}$ is utility of a system, $\phi$ is a function on n-dimensional
vector of all system properties $K_{S}^{n}=[k_{1},k_{2},...k_{n}]$
given event $e_{i}$, and $P(e_{i})$ is the probability (or relative
frequency), of event $e_{i}$.

The definition is universal because it bases value on a scenario.
Utility of same system in different scenarios differs (and may be
even negative), but it always deterministically follows from system
properties.

Function $\phi$ calculates system utility in case of given event.
In order to support claim that network has size-dependent value, we
need to show that size-dependent component of $\phi$ can be separated
from event-dependent one. Even when we compare systems that differ
in only one parameter, we cannot extract event-independent component
because $a(x,y)=b(x)c(y)$ has no solutions. It shows that network
has no universal size-dependent value. 

But it is plausible to assume that there exists a non-empty set of
events $E$ for which we can represent $\phi$ as $\phi=\xi(e_{i})\psi(k)$,
where $\xi(e_{i})$ and $\psi(k)$ are event-dependent and property-dependent
components of $\phi$. Note that $\psi(k)$ is independent of any
event, and therefore total utility can be represented as
\[
U=\psi(k)\sum_{i=1}^{\infty}\xi(e_{i})P(e_{i})
\]
 In this case, if systems differ only in $k$, total utility $U$
can be represented as $U=\psi(k)C$, where $C$ is a system-independent
factor. External factors can influence $C$, but as long as our system
has only one varying parameter, its utility is proportional to a function
of that parameter. Now we need to make sure our model meets that criterion.

 Information Network is a collection of information consumers and
producers connected by communication channels.
\end{defn}
In order to make size the only property that distinguishes two networks,
we have to add that all nodes and channels are identical, and also
that number of channels is a function of network size. Let's also
discard constraints by assuming that all parts of a network can process
infinite amount of data in no time. Those assumptions are enough to
make the notion of network value for set of events $E$ mathematically
consistent. In the next chapters I study and improve this model.

\section{Network Effect}

Common\cite{wikineteffect} understanding of network effect is both
simple and compelling. The primary function of a network is connecting
users. Therefore the value of a network to one user is proportional
to the number of other users: $Q\propto N-1$ ($\propto$ denotes
proportionality). As a result, total value of a network is proportional
to $N(N-1)=N^{2}-N$. Note that it is identical to the maximum number
of unique directed links between nodes.

Let's analyze that claim. By viewing many to many communication as
a simultaneous mutual broadcasting, we can model the network as a
superposition of broadcast-type networks. See figure 1.

\begin{figure}[tbh]
\begin{centering}
\textsf{\includegraphics[scale=0.165]{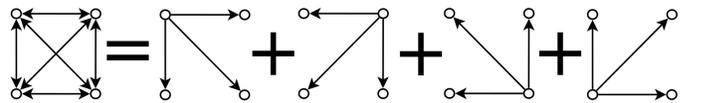}}
\par\end{centering}

\caption{Switched network of size $N$ can be represented as superposition
of $N$ virtual broadcast networks with $N-1$ receivers in each}
\end{figure}

Network effect basically assumes that those networks are independent.
However closer investigation suggests that they are not: though all
virtual transmitters indeed are independent, they share same receivers.
As a result, links compete for nodes, and each link's share of its
terminals equals $\frac{1}{N-1}$. Consequently, in a network with
higher density of links one link has proportionally less value. For
example, when above network expands $X$ times, each link's share
of its terminals decreases $X$ times. To illustrate this with telephone
networks, potential to place a call for each phone increases, but
potential that the call will be answered decreases proportionally.
In other words, the gain is imaginary.

One can argue that sharing is applicable only when node capacity is
severely constrained. But that is not the case. We need $N-1$ links
to make a network of size $N$, and other links are redundant. That
is why net value of a network is always $N$. Network effect suggests
tendency to exponential growth of partially redundant data, not of
value.
\begin{lem}
Direct link in unconstrained network has no value. \end{lem}
\begin{IEEEproof}
In zero-latency, infinite bandwidth network of size $N$, node $X$
has direct connections to every other node. That results in $N-1$
direct links. Node $Y$ has just one connection (of course, to $X$).
As a result, we obtain two subnetworks. Network $A$ has $N-1$ nodes
(all except $Y$), and network $B$ has $2$ nodes: $X$ and $Y$.
Next, we set $X$ to bridge $A$ and $B$. Now $Y$ gets exactly the
same network service as $X$, and the only solution of $V_{lnk}\times(N-1)=V_{lnk}\times1$
is $V_{lnk}=0$.\end{IEEEproof}
\begin{cor}
From 1 follows that all network value is contained in nodes. 
\end{cor}
This hints us that information network is identical to a non-distributed
information system, such as the computer. In computer memory, you
always get same amount of memory from $N$ units regardless of how
you connect those units. Notion that a collection of interlinked nodes
can enjoy a non-linear increase in value is mathematically inconsistent,
and there are many ways to prove it. For example we can split every
network node in $n$ parts and connect those parts back to the network,
which, according to network effect, must raise its value $\frac{n^{2}}{n}=n$
times, which is contradiction.

\section{Network Efficacy}

Describing network effect as exponential growth of value might originate
from difficulties to discriminate redundant data from valuable information.
However that neither proves nor disproves Metcalfe's law. The thing
is, describing network as a collection of nodes is not the best way
to model it. A better point of view is that network has value in previous
sense. It provides connectivity between certain parts of (potentially
larger), physical or business system, and it is that larger system
that has value. Here are the corresponding definitions:
\begin{defn}[Information System]
 The information system is a collection of complementary subsystems
(nodes), that contribute to overall information value.
\begin{defn}[Telecommunications Network]
 The telecommunications network is an apparatus for information transmission
between flexible number of terminals.
\end{defn}
\end{defn}
According to later definitions, the telecommunications network is
just one of possible transmission agents for an information system;
it differs from the system itself. As an illustration, the broadcast
network can be viewed as a complex business that delivers its products
via telecommunications network. According to the definition, Information
System is too complex to derive value from its size. Because of that,
information network is a poor approximation of an information system.

To replace value with something more appropriate for a network, let
me introduce another definition. It is the efficacy of a network.
\begin{defn}[Network Efficacy]
 Network Efficacy is the amount of useful data an underlying information
system is able to send through a network of $N$ identical nodes.
It is the product of network size and node communication efficacy
$\zeta$:
\[
\psi=N\zeta
\]

As before, we investigate an unconstrained network with identical
nodes that limits $\zeta$ neither by bandwidth nor by node capacity.
Note that Metcalfe's law holds only when $\zeta$ is proportional
to $N$. If $\zeta$ is constant, network properties are linear in
$N$. As we will see in a moment, $\zeta$ exhibits both of those
behaviors. Now let me describe a model that reveals this. 
\end{defn}

\section{Deficit Model}

Consider the following scenario: User $X$ wants to contact certain
members of her family over Skype. These people can be described by
set $A\subseteq\Omega$, where $\Omega$ denotes all nodes of the
information system that corresponds to $X$'s family. 

Skype as a network represents an independent set $B$. Those of $\Omega$
members who use Skype comprise the effective network $E=B\cap\Omega$.
Note that the actual size of the Skype network is irrelevant Effective
network size equals $|E|$ ($|...|$ denotes cardinality). $X$ can
contact only those people that belong to $A\cap E$ (the intersection
of her contact list and her relatives in Skype network). Expected
proportion of people from $A$ that Skype allows to contact is independent
of $A$. It is $\frac{|E|}{|\Omega|}$, or effective network size
divided by information system size. Note that $|E|\le|\Omega|$ because
$E\subseteq\Omega$. The same is true for every member of $E$, which
leads to the following:
\begin{thm}[Network Efficacy Theorem]
 Network Efficacy is proportional to square of effective network
size divided by information system size, where effective network represents
the networked part of the information system. 
\[
\psi=\alpha\frac{N_{E}^{2}}{N_{\Omega}}
\]
where $\alpha$ is size-independent transmission rate, $N_{E}=|E|=|B\cap\Omega|$,
and $N_{\Omega}=|\Omega|$.\end{thm}
\begin{IEEEproof}
To avoid projections to the future, let us investigate what happens
when a network suddenly disconnects part of its nodes: $N_{E}=N_{\Omega}/x$
(1). The disconnected nodes try to contact the network at a constant
rate $\alpha$ and receive errors. Remaining nodes also contact old
address space, therefore their success rate is proportional to remaining
fraction of the network: $\psi=\alpha\frac{N_{E}}{x}$ (2). From 1,
$x=\frac{N_{\Omega}}{N_{E}}$ (3). Substituting 3 into 2 we get $\psi=\frac{\alpha N_{E}^{2}}{N_{\Omega}}$.
The cause of non-serviceable requests is that active information system
is larger than its accessible part. Therefore, regardless of whether
network shrinks, grows, or stays constant, average fraction of satisfied
demand is proportional to the square of networked fraction of the
information system.
\end{IEEEproof}
The next section illustrates that by example.

\section{Heterogeneous Networks}

Deploying a network inside of an information system implies that there
exists an old way of communication between links. In fact, telecommunication
networks constantly replace one another, and most networks are heterogeneous.
\begin{cor}
From Network Efficacy Theorem follows that when there is a default
network $D$ that connects all nodes and a preferred network $K$
that connects part of those nodes, 
\[
\psi_{K}+\psi_{D}=\frac{1}{1-n^{2}}
\]

where $\psi_{D,}+\phi_{K}$ denotes the joint capacity of default
and preferred network, and $0\leq n<1$ is the fraction of nodes that
the preferred network connects.
\end{cor}
Let me illustrate above notion by example: Parallel computation cluster
nodes produce synchronous traffic at a constant rate. All traffic
is produced and consumed locally; it is evenly distributed among nodes.
Nodes are connected by a network switch with total flow cap of $1$
Tb/s. An additional, separate network with flow capacity of $2$ Tb/s
is being deployed. The task is to fully load both of networks. When
two networks depend on each other, total traffic is limited to $\frac{1}{1-n}$
of the smaller capacity, where $n$ is fraction of traffic that goes
over the faster network. If faster network took $\frac{2}{3}$ of
the load from the slower one, total capacity would be $3$ Tb/s. As
far, workers have installed faster network on $\frac{2}{3}$ of the
nodes. What is the overall capacity of the network?

\textbf{A}: Efficacy of the faster network $\psi_{F}=\frac{(2/3N_{\Omega})^{2}}{N_{\Omega}}=\frac{4}{9}$
of the slower network. As was already shown, it happens because nodes
connected by the faster network can find a destination inside it only
$\frac{2}{3}$ of the time, therefore $\frac{1}{3}$ of time they
default to the slower network. As a result, total network capacity
is $1/(1-n^{2})=1.8$ Tb/s. To achieve $3$ Tb/s, workers need to
connect $n=\sqrt{1-\frac{1}{\psi_{D}+\psi_{K}}}\approx81.65$\% of
nodes.

\section{Saturated Behavior}

When network size equals to information system size, adding new nodes
results in equal growth of $N_{E}$ and $N_{\Omega}$. As a result,
$\zeta$ stays constant. See figure 2.

Though each node accesses increased address space, each address has
proportionally less value. The same applies to the amount of obtainable
information.

\begin{figure}[tbh]
\includegraphics{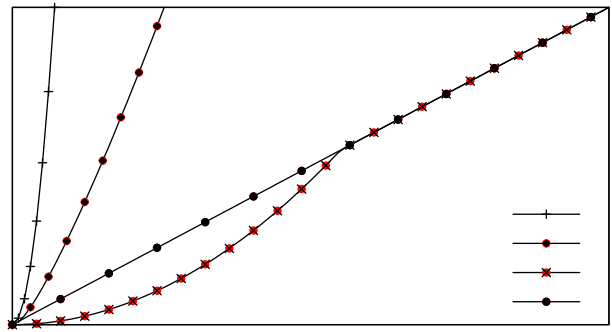}

\caption{After a network catches the size of its underlying information system,
it switches to linear growth}
\end{figure}

\section{Multipurpose Networks}

A multipurpose network is one that serves many information systems.
Total efficacy of a multipurpose network is the sum of its capacities
for all systems. Therefore, overall size of a network may serve as
a rough approximation of its efficacy. 
\[
\psi_{mp}^{tot}=\sum_{i}\psi_{i}\propto N^{2}
\]

\section{Discussion}

Studied models did not confirm moderate exponential growth estimates
suggested by \cite{briscoe09f:nlogn-icr}. Their analysis also invalidates
a notion that network effect can lead to exponential growth of utility.
Instead, it suggests that when a network grows by extending existing
or creating new information systems, its value grows linearly, regardless
of the number of spawned systems. One explanation is that simple network
model is just a distorted approximation of a distributed information
system, and as such it has dynamics that is common to information
systems.

However in a more complex setup, when a network grows within a larger
information system, its efficacy raises exponentially until it catches
the size of the underlying information system. That exponential growth
exactly follows Metcalfe's law. Efficacy dramatically affects network
utility, and may play a major role in limiting network value in heterogeneous
environments.

In addition to linear growth of information, model suggests linear
growth of redundancy. As information density is a reciprocal of data
redundancy, network effect dictates that unique content of overall
network traffic is a reciprocal of network size. However the later
effect may be mitigated by limited applicability of network model
to real-world information systems. It can be argued that information
systems built with network model in mind, such as WWW, exhibit more
redundancy than setups based on complementary subsystems. That suggests
promotion of node complementarity as a possible way to reduce redundancy.

\appendices{}

\bibliographystyle{plain}
\bibliography{spectrum}

\end{document}